# EXCHANGE PARAMETERS IN THE TCNE-BASED MAGNETS AS ESTMATED WITH USE OF THE EFFECTIVE HAMILTONIAN CRYSTAL FIELD METHOD

A.L. Tchougréeff

Poncelet Laboratory, Independent University of Moscow,
Bol. Vlasyevskii per., 11, Moscow, 119002, RUSSIA

**Abstract:** Estimates of the exchange interaction parameter ($J$) between electronic magnetic momenta residing in the d-shells of several divalent first row transition metal ions, and those residing on the $[TCNE]^-$ anion-radicals (TCNE = tetracyanoethylene) that are bonded to the metal ion are obtained by the use of the EHCF method in the cluster approximation. The clusters have common composition of $[M^{II}(NCMe)_5TCNE]^+$. Their geometry was set according to that observed for $Mn^{II}(TCNE)_{1/2}(I_3)_{1/2}$ and $M^{II}(TCNE)(C_4(CN)_8)_{1/2}$ (M = Fe, Mn) with the $[TCNE]^-$. For the $Mn^{II}(TCNE)_{1/2}(I_3)_{1/2}$ two cluster types are considered: one with the geometry corresponding to $[TCNE]^-$ in the two-dimensional layer, and another connecting the layers. For $M^{II}(TCNE)(C_4(CN)_8)_{1/2}$ only the layer position of the $[TCNE]^-$ is considered. In all cases the ground state spin of the d-shells as coming from the (EHCF) calculation is high-spin in agreement with experiment. Also, in all cases the LUMO (singly filled in the material) is predominantly concentrated on the ethylenic C-C bond and is antibonding. The $J$ value for the $Mn^{II}(TCNE)_{1/2}(I_3)_{1/2}$ layer is at least as twice as strong as one for the similar geometry for $M^{II}(TCNE)(C_4(CN)_8)_{1/2}$. The correct order of magnitude of the calculated exchange parameters, as compared with those extracted from experiments with use of the mean-field approximation indicates the general validity of this approach. However, the direct comparison of these $J$ values with the critical temperatures is not possible since these latter heavily depend on details of the interlayer coupling not available through the EHCF procedure. In order to get more information on the geometry dependence of the exchange parameters, the dependence on the tilt angle of the TCNE with respect to the plane formed by the metal ions and four equatorial NCMe groups was computationally investigated. In all cases, a remarkably strong dependence, which, however, has different trends depending on the number of electrons in the *d*-shells of respective transition metal ions was observed.

## *Introduction*

Organic-based magnets based on the $[TCNE]^{-\cdot}$ (TCNE = tetracyanoethylene) radical anion have magnetic ordering temperatures ($T_c$) as high as 400 K.[1,2,3]. Examples include zero-dimensional (0-D) [Fe(C$_5$Me$_5$)$_2$][TCNE] [4,5], 1-D [MnTPP][TCNE] (H$_2$TPP = *meso*-tetraphenylporphyrin) [6,7], 2-D [Fe(TCNE)(NCMe)$_2$][FeCl$_4$] [8] and M(TCNE)[C$_4$(CN)$_8$]$_{1/2}$ (M = Mn [9,10], Fe [8,11]), and 3-D Mn$^{II}$(TCNE)$_{1/2}$(I$_3$)$_{1/2}$ [9]. V(TCNE)$_x$ (x ~2) has the highest $T_c$ ( ~ 127 $^o$C), but is amorphous and not structurally characterized, but is anticipated to have a 3-D structure [12].

The strong nearest-neighbor magnetic coupling than leads to magnetic ordering is attributed to direct exchange between the $S$ = 1/2 [TCNE]$^{\bullet-}$ and the $S$ > 1/2 metal ion to which it is bonded. Each [TCNE]$^{\bullet-}$ bonds to two or four metal ions, and each metal ion bonds to 2, 4, or 6 [TCNE]$^{\bullet-}$'s except for 0-D materials, and form an extended strongly magnetically coupled network. To understand the type and magnitude of the nearest neighbor exchange, $J$, was computationally evaluated as a function of the metal ion it was bonded to as well as ∠M-N-C$_{TCNE}$.

The [M$^{II}$(NCMe)$_5$(TCNE)]$^+$ moiety was used to computationally study $J$, as M$^{II}$ is surrounded by six nitriles, one of which is a TCNE. Since $\mu_4$-[TCNE]$^{\bullet-}$ is present, the key structural parameter is the ∠M-N-C, which varies from 159.1 to 171.5$^o$ for Mn$^{II}$(TCNE)$_{1/2}$(I$_3$)$_{1/2}$ [9] for M$^{II}$ = V, Cr, Mn, Fe, Co Ni, and Cu.

## *Theoretical method*

### Effective Hamiltonian Crystal Field method

The Effective Hamiltonian Crystal Field (EHCF) method is a direct semi-empirical calculation of the crystal field felt by the *d*-shell of a transition-metal ion in a complex or solid on the basis of composition and geometry with a minimal fit of parameters used [13]. This is reached by assuming the many-electron wave function of a complex in the form:

$$\Psi = \Phi_d(n_d) \wedge \Phi_L(n_L), (1)$$

where $\Phi_d$ is the full configuration-interaction function of $n_d$ electrons in the *d*-shell of the transition-metal ion and $\Phi_L$ is the function of all other electrons ($n_L$) of the system taken in a semi-empirical self-consistent field approximation. The symbol ∧ refers to the fact that the above "product" function is antisymmetric (*i.e.*, changes its sign) when the coordinates of each pair of $n_d+n_l$ electrons are interchanged. This type of the wave function formalizes the usual assumptions of the crystal field theory, but it allows for calculation of the crystal field. That latter is contributed by the ionic and covalent parts: $W^{ion}_{\mu\nu}$ and $W^{cov}_{\mu\nu}$ given, respectively, by the formulae:

$$W_{\mu\nu}^{ion} = \sum_{L} V_{\mu\nu}^{L} Q_L + \sum_{i \in 4s, 4p} g_{\mu i} P_{ii} \quad (2)$$

($Q_L$ is the effective charge of the $L^{th}$ ligand of the complex, $P_{ii}$ is the diagonal element of the one-electron density matrix over the metal 4s- and 4p-AOs all obtained from the wave function $\Phi_L$; $V_{\mu\nu}^{L}$ are matrix elements of the electrostatic crystal field induced by a unit charge located in the position of the $L^{th}$ ligand of the complex, $g_{\mu i}$ is the parameter of the Coulomb interaction between electrons in the $\mu^{th}$ d-AO and the $i^{th}$ 4s- or 4p-AO of the metal atom); and

$$W_{\mu\nu}^{cov} = \sum_{j \in MO} \beta_{\mu j} \beta_{\nu j} \left( \frac{n_j/2}{\Delta E_{j \to d}} - \frac{1 - n_j/2}{\Delta E_{d \to j}} \right), \quad (3)$$

where $n_j$ is the occupation number of the $j^{th}$ ligand MO in the single-determinant wave function $\Phi_L$ and is equal to 2 or 0, $\beta_{\mu i}$ is the one-electron hopping integral between the $\mu^{th}$ d-AO and $j^{th}$ ligand MO, $\Delta E_{j \to d}/\Delta E_{d \to j}$ are the energies of the states with one electron transferred between the d-shell and the $j^{th}$ ligand MO, calculated according to the formulae:

$$\beta_{\mu i} = \sum_{l} c_{il} \beta_{\mu l}$$

$$\beta_{\mu l} = \beta^{ML} (\beta_d + \beta_l) S_{\mu l}$$
$$\Delta E_{j \to d} = I_j - A_d - g_{dj} \quad , (4)$$
$$\Delta E_{d \to j} = I_d - A_j - g_{dj}$$

where $A_j$ and $I_j$ are, respectively, the electron affinity and the ionization potential corresponding to the $j^{th}$ MO of the l-system; $A_d$ and $I_d$ are, respectively, the electron affinity and the ionization potential of the d-shell estimated as free-ion values shifted by the Coulomb potential of the effective charges in the ligands, and the $c_{il}$'s are the MO LCAO coefficients taken from the calculation for the wave function $\Phi_L$. The quantities $\beta_d$ and $\beta_l$ are the hopping parameters characteristic for the given transition-metal atom and given donor atom separately; $S_{\mu l}$ are the overlap integrals between the $\mu^{th}$ d-orbital of the metal ion and the $l^{th}$ AO of the ligands. The parameter $\beta^{ML}$ – specific for a pair of the transition-metal atom and the donor atom – scales the one-electron hopping $\beta_{\mu l}$ integrals between the d-shell and the orbitals $l$ located on the donor atoms. The quantities $g_{dj}$ are the energies of the Coulomb interaction between electron and hole located in the d-shell and the $j^{th}$ ligand MO.

It has been shown that the hopping integrals together with the energies of the states with electrons transferred to/from the *d*-shell and from/to the ligands ("charge-transfer states") control up to 90% of the observed splitting [14-24]; the rest is given by the Coulomb field of the effective charges in the ligands. These quantities are uniquely defined by the chemical composition and by the molecular spatial structure. This specificity is explicitly reflected by the values of the MO-LCAO coefficients of the *l*-system $c_{il}$, of the orbital energies $\varepsilon_j$ of the latter, as well by the (geometry-dependent) overlap integrals $S_{\mu l}$, quantities $\beta_d$ and $\beta_l$ having the dimension of energy, and the scaling factor $\beta^{ML}$ together defining the hopping matrix elements characteristic for the given transition-metal atom and given ligands. Applying the EHCF method to a variety of systems both in molecular [14-18, 20, 21, 23, 24] and crystalline (in the cluster approximation) [19,22, 25-27] settings have shown an unprecedented precision of the estimates of the amount of the crystal fields and thus of the d-d excitation spectra. In the following Section we describe how this method can be used for estimating the exchange parameters in the TCNE based magnets.

## Using EHCF for estimating exchange parameters

The exchange interactions for organic-based magnets occur between the local electronic momenta of different *d*-shells are mediated by diamagnetic "organic" groups or ions [28]. The magnitude of the effective exchange parameters quantifying these interactions do not usually exceed several dozens of meV and are estimated by the expressions of the form:

$$\sim \frac{\beta_{dl}^4}{\Delta E_{dl}^2 \Delta E_{dd}}, (5)$$

where $\beta_{dl}$ is the parameter describing one-electron hopping between the states of the d-shell and the states in the "organic" ligands, $\Delta E_{dl}$ is the energy of a charge transfer state with an electron (hole) transferred from a d-shell to a ligand state, and $\Delta E_{dd}$ is the energy of a state with one electron transferred between different d-shells in the material. Strong angular dependence of the resonance integrals $\beta_{dl}$ is the basis of the well-known empirical Goodenough-Kanamori rules [28] However, comparing the expression eq. (3) for the dominant – covalent contribution to the crystal field felt by a *d*-shell in the ligand environment with the estimate eq. (5) on one hand and the magnitudes of the crystal field splitting with those of the characteristic exchange parameters on the other one can easily see that the estimates of the exchange parameters are two order of magnitude smaller and appear in two orders higher in $\beta_{dl}$ than the crystal field splitting. Respectively the precision requirements for the exchange parameters are in general case two orders of magnitude higher than for the crystal fields.

For the M-TCNE magnets the local states responsible for the formation of the magnetic structure are not exclusively the transition metal $d$-shells, but include the singly occupied MO's (SOMO's) of the "organic" TCNE units. Correspondingly the underlying physics behind this structure is precisely the one-electron hopping between the SOMO of the [TCNE]$^{\bullet-}$, and the orbitals that are singly occupied in the ground states of the respective $d$-shells. Thus, the exchange $J$ controlling the low-temperature magnetic behavior of these materials (and determining the range of "lower" temperatures) are those between the singly occupied $d$-orbitals and the SOMO's of the [TCNE]$^{\bullet-}$ :

$$J = \frac{1}{n_{unp}} \sum_\mu J_\mu$$

$$J_\mu = 2\beta^2_{\mu,SOMO} \left( \frac{1}{\Delta E_{d \to SOMO}} + \frac{1}{\Delta E_{SOMO \to d}} \right), \quad (6)$$

where the sum over $\mu$ is extended to the singly occupied $d$-orbitals, and $n_{unp}$ is the number of unpaired electrons in the d-shell (or equivalently the number of terms in the sum over μ). Comparing the eq. (6) with that for the covalent contribution to the crystal field eq. (3) we note that the EHCF method requires the same quantities to calculate the leading contribution to the crystal field as those needed to estimate $J$. Indeed the EHCF program suit [13] routinely calculates the required quantities, however, the EHCF package imposes certain restrictions working around which will be described below.

## Tests and workaround the EHCF limitations

The EHCF method was originally developed for modeling the $d$-$d$ excitation spectra in the isolated molecules of the transition metal coordination compounds. With certain precautions, it can be used for modeling transition metal ions in the solid state environment (specifically, the cluster approximation is to be used with a reasonably sized and correctly defined clusters). Further problem is that the EHCF method implemented so far requires the ligand system to have the closed electronic shell [with all ligand MO's doubly filled – note the allowed values of the MO occupation numbers right below eq. (3)]. Thus, the results obtained from presently workable models must be adapted in such a way that they become valid for solids containing unpaired electron in the $l$-system. Specifically, [M$^{II}$(NCMe)$_5$TCNE]$^+$ was used for the calculations, and the number of the $d$-electrons corresponds to the electronic structure of M$^{II}$. Five N≡C-C units originating either from TCNE or their $[C_4(CN)_8]^{2-}$ dimers are terminated by hydrogen atoms, thus yielding the model NCMe ligands. NCMe is a common solvent and many $[M^{II}(NCMe)_x]^{2+}$ complexes, all high spin, have been structurally characterized, and are good models for MII surrounded by six [TCNE]$^{\bullet-}$ ions. We performed series of calculations of the ligand subsystems of these models.

The ground states of the respective *d*-shells for a series of geometries with the varied bending of the TCNE ligand with respect to the base of the M(NCMe)$_5$ unit were computed. The ∠MNC was varied by $60^{\circ}$. The calculations show that the model LUMO is well localized on the TCNE, and is of π-symmetry with respect to the TCNE plane at all intermediate geometries. It is approximately antisymmetric with respect to the reflection in the mirror plane perpendicular to the ethylenic C-C bond and is always ethylenic C-C antibonding. These results enable the identification of the LUMO of the [M(NCMe)$_5$TCNE]$^+$ modelwith the SOMO of the TCNE. Moreover, the LUMO's of the [M(NCMe)$_5$TCNE]$^+$ are localized in terms of the contribution of the AO's located on atoms of the model other than those of the TCNE unit and (2) is isolated on the energy scale being separated by more than 10 eV from the occupied orbitals of the model and by more than 3 eV from the higher unoccupied model orbitals that in their turn form a dense group of states separated by tenths of eV. These features are fairly the assumptions of the known Koopmans' theorem. which is based on the invariance of the shape and energy of an orbital under adding or withdrawing one electron to or from the system. It indirectly justifies our model containing only one TCNE unit, since including a larger number of them would cause an appearance of numerous almost degenerate SOMO's that could in their turn become hybridized with each other. Following the above identification of the model LUMO with the material's SOMO allows us to identify the resonance integrals of the d-orbitals of the metal atoms and the model LUMO's available from the EHCF suit output with those required in eq. (6): $\beta_{\mu,SOMO} = \beta_{\mu,LUMO}$. This approximation is reasonable for the purpose of studying qualitative geometry dependence of the exchange parameters since the latter is predominantly controlled by that of the squared resonance integrals $\beta^2_{\mu,SOMO}$. However, in order to operate with the quantities with the dimensionality of energy the squared resonance integrals are divided by the energies of the charge transfer states $\Delta E_{d \to LUMO}$ as well available from the EHCF suit and sum the contributions coming from different singly occupied *d*-orbitals. Although, these energies are not those required in the energy denominators of eq. (6) they have the same order of magnitude and weakly depend on geometry. These results are given in Table 2 and depicted in Figure 2.

In order to obtain more realistic estimates of *J* for the [TCNE]$^{\bullet-}$ as present in the TCNE-

based materials we have to consider the energy denominators in more detail. Like the estimates of the resonance integrals they have to be derived from the results obtained from the EHCF program suit for the $[M(NCMe)_5TCNE]^{2+}$ model. As noted

$$\Delta E_{d \to SOMO} = I_d - A_a - g_{dSOMO}$$

$$\Delta E_{LUMO \to d} = I_a - A_d - g_{dSOMO}$$ (7)

In eq. (7) the ionization potentials, electron affinities, and the interaction integrals must be those for the structures under study. From our model EHCF calculation we can easily obtain similar values for the divalent metal ions surrounded by the neutral NCMe and TCNE molecules:

$$\Delta E^{(0)}_{d \to LUMO} = I^{(0)}_d - A^{(0)}_a - g_{dLUMO}$$

$$\Delta E^{(0)}_{LUMO \to d} = I^{(0)}_a - A^{(0)}_d - g_{dLUMO}$$ (8)

The ionization potentials and electron affinities of the d-shells in the crystals and in the models (with the zero superscript) are given by respectively by the expressions:

$$I^{(0)} = -U_{dd} - (n_d - 1)g_{dd} - V_{Field}(TCNE)$$
$$A^{(0)}_d = -U_{dd} - n_d g_{dd} - V_{Field}(TCNE)$$
$$I_d = -U_{dd} - (n_d - 1)g_{dd} - V_{Field}(TCNE^{\bullet -})$$
$$A_d = -U_{dd} - n_d g_{dd} - V_{Field}(TCNE^{\bullet -})$$
$$A^{(0)}_a = -\varepsilon_{LUMO}$$
$$A_a = -\varepsilon_{LUMO} - g_{aa}$$ (9)

The calculations on the model molecules $[M(NCMe)_5TCNE]^+$ yield the quantities with the (0) superscripts. In order to improve our estimates of the exchange in real materials we need the respective quantities without superscripts. Since by assumption the LUMO shape changes insignificantly when going from neutral TCNE to $[TCNE]^{\bullet -}$ (where it becomes the SOMO), we can estimate $g_{dSOMO} = g_{dLUMO}$. Under these conditions the only effect of the one-electron reduction of TCNE is the difference of the respective contributions to the potential energy of the d-electrons as coming from the radical-anion ($V_{Field}(TCN^{\bullet -})$) and the neutral TCNE units ($V_{Field}(TCN)$). This difference is, however, easily estimated:

$$g_{dSOMO} = V_{Field}(TCNE^{\bullet -}) - V_{Field}(TCNE)$$ (10)

The value of $g_{dSOMO}$ is thus easily found from the values of the ionization potential of the d-shell, the orbital energy of the LUMO εLUMO and the energy of the metal to ligand charge transfer state $\Delta E_{d \to LUMO}^{(0)}$ that are calculated by the EHCF package. We collect in Table 1 these and other quantities required for the definite calculation of the exchange integrals as extracted from our EHCF calculations. The only missing value is that of the energy of the electron-electron repulsion for the SOMO $g_{aa}$ required to estimate the difference between the ionization potential and electron affinitiy of the SOMO (going in this respect beyond the Koopmans' one-electronic picture). We estimate it to be of ca. 1 eV and to be geometry independent. With these simple assumptions all elements of the construction are available.

## *Results and Discussion*

### Geometry dependence of exchange energy $E^0_{ex}$

The geometry dependence of the exchange energy $E^0_{ex}$ characteristic for each transition metal ion in the model acetonitrile/TCNE environment has been computed via eq. (6), and the energy $\Delta E_{d \to LUMO}^{(0)}$ of the *d*-shell to LUMO charge transfer state as a characteristic energy denominator. The ∠MNC was decreased by up to 60° departing from the respective experimental geometries (for the Mn, Fe, and Co compounds) with a step of 15°. For the room-temperature V-TCNE magnet the structure is not known for sure experimentally until now so that we used the structure [30] obtained by the numerical optimization. For the Cu, Ni, and Cr compounds no structural data are available. Thus we adjusted the M-N distances used for the Co compound according to the differences between the Shannon ionic radii of these cations (hexacoordinated, high-spin) [31].

The parameters used in calculations by the EHCF procedure for Cu, Ni, Co, and Fe had been taken standard from papers; those for Mn – from paper. Nitrogen-containing compounds of V and Cr had not been calculated by the ECHF so far. Thus we estimated the resonance scaling parameters $\beta^{ML}$ for $M$ = V, Cr and $L$ = N to be respectively 0.975 and 1.03 in order to reproduce the $10Dq$ values [32] for the respective $[M(NCMe)_6]^{2+}$ cations with the *M*-N distances in the similar compounds [33].

The exchange energies are strongly geometry dependent as is the covalent contribution to the *d*-shell splitting eq. (3). Similarly to that contribution the exchange energies depend on molecular geometry through the overlap integrals exponentially varying with interatomic separations and depending on angles between the orbitals involved. Bending of the ∠MNC significantly increases the overlap between the $d_z^2$- and $d_{x^2-y^2}$-orbitals (we assume the model cluster geometries in which the TCNE nitrogen coordinated to the metal ion takes the position on the *x*-axis) and the TCNE LUMO (of π-symmetry with respect to the TCNE plane). At the zero bending – the ∠MNC angle of ca. 180° – their overlap is almost vanishing (it would be precisely vanishing at 180°), whereas at the 60° bending of the layer geometry of the Mn(TCNE)$_{1/2}$(I$_3$)$_{1/2}$ compound brings the TCNE unit in a position where its plane is almost parallel to the base of the M(NCMe)$_5$ unit corresponds to the value of the ∠MNC of ca. 90° (at 90° the overlap with the $d_z^2$-orbital would be maximal), whereas other overlaps decrease. First of all it applies to that of the $d_{xz}$-orbital with the TCNE LUMO: in the layer geometry at zero bending this overlap is maximal as one between the orbitals with π-symmetry with respect to the TCNE plane. Under bending the ideal alignment of these π-orbitals is broken and the corresponding overlap decreases. (Another difference between the wall and layer geometries is the direction of the bending: in the wall geometry the bending is such that the TCNE plane remains parallel to the basal N-Mn-N axis of the Mn(NCMe)$_5$ unit whereas in the layer geometry the TCNE plane is parallel to one of the diagonals of the equatorial MnN$_4$ square).

Irrespective to the minor details of the geometry the different trend of the exchange energy as a function of the bending angle in case of Mn, Fe, Co, and Ni compounds, and the V compound is remarkable. In case of Mn, Fe, Co and Ni compounds the $d_z^2$- and $d_{x^2-y^2}$-orbitals are singly occupied and thus contribute to the total exchange of the *d*-shells with the SOMO of the [TCNE]$^{•-}$. Their increase with bending is responsible for that of the exchange energy. By contrast, in the V compound the $d_z^2$- and $d_{x^2-y^2}$-orbitals remains empty for all geometries, and thus increase of its overlap with the SOMO does not contribute to that of the exchange, which decreases with increase of the bending since the overlaps of the involved *d*-orbitals decrease. The Co compound is remarkable by its nonmonotonous variation of the exchange energy with the bending. In this compound only one of the $d_z^2$- or $d_{x^2-y^2}$-orbitals is singly occupied. At smaller bending angles the $d_z^2$-orbital is singly occupied whereas the $d_{x^2-y^2}$-orbital is empty. The overlap of the TCNE SOMO with the $d_z^2$-orbital rapidly increases thus yielding respective increase of the exchange energy. However, at the 60° bending the occupancy of the $d_z^2$- or $d_{x^2-y^2}$-orbitals is interchanged and now the interaction of the $d_{x^2-y^2}$-orbital with the TCNE SOMO contributes to the overall exchange energy. The overlap of the $d_{x^2-y^2}$-orbital with the TCNE SOMO, although increases with the bending, remains always smaller than that of the $d_z^2$-orbital, which explains the obtained nonmonotonous run of the exchange energy in the Co model with the bending.

The Cu model is somewhat isolated among others considered in this Section. Like in all other cases the bending of the TCNE unit results in significant increase of the overlap between the TCNE SOMO and the $d_z^2$-orbital. The latter, however, remains doubly occupied for all considered geometries as one could expect. The singly occupied orbital in the $Cu^{2+}$ ions which is only possibly involved in magnetic exchange is the $d_{x^2-y^2}$-orbital. Its overlap with the TCNE SOMO although increasing with the bending angle remains not that large as the results of Table 2 show. In this situation one can expect even *ferromagnetic* overall exchange due to other (superexchange) mechanisms involving strongly overlapping doubly occupied $d_z^2$-orbital may enter into play at least at smaller bending.

## Refined exchange parameters for realistic geometries

The above procedure for refining the values of the energy denominators (the energies of the charge-transfer states involving the SOMO in the TCNE based solids) had been applied to the cluster models based upon experimentally available structures of the corresponding Mn and Fe based materials. For V(TCNE)$_2$ the experimental structure is not known so that we used the VASP optimized structure [30]. The results of so performed calculations for the clusters are given in Table 3.

The largest exchange integral estimated according to the proposed procedure belongs to the material manifesting the highest observable critical temperature of the magnetic ordering in the family of the TCNE-based magnets, namely, for of V(TCNE)$_x$. Although this value is still too high as compared with the experimental estimates based on the spin-wave picture of the magnetic phase diagram of HM for V(TCNE)$_2$ [30] it is much better than that extracted from the DFT based calculation [34]. One has to realise that only the dominating contribution to the *d*-shell – TCNE SOMO exchange – the kinetic antiferromagnetic one – has been taken into account in the present study. Although, further contributions to this quantity are estimated to be an order of magnitude smaller and to have opposite sign [28] their inclusion will be necessary to assure correct numerical values. As for other materials listed in Table 3 their respective calculated exchange parameters are in fair correlation with the order of the corresponding characteristic temperatures.

*Acknowledgments*

This work is supported by the RFBR through the grant No 10-03-00155. The author is grateful to Prof. J. S. Miller (University of Utah) for valuable discussion and sending structural data on the compounds considered in this paper. Parts of this work have been performed during author's stay in the Institute of inorganic chemistry RWTH – Aachen University, whose hospitality is gratefully acknowledged.

*References*

[1] V. I. Ovcharenko, R. Z. Sagdeev. *Russ. Chem. Rev.* **1999**, *68*, 345; O. Kahn. *Adv. Inorg. Chem.* **1995**, *43*, 179; D. Gatteschi. *Adv. Mat.* **1994**, *6,* 635; J. A. Crayson, J. N. Devine, J. C. Walton. *Tetrahedron* **2000**, *56*, 7829; S. J. Blundell, F. L. Pratt. *J. Phys.: Condens. Matter*, **2004**, *16*, R771.

[2] J. S. Miller, A. J. Epstein. *Angew. Chem. Int. Ed. Engl.* **1994**, *33*, 385.

[3] J. S. Miller, *Chem. Soc. Rev.* **2011,** *40*, to appear.

[4] J. S. Miller, J. C. Calabrese, A. J. Epstein, R. W. Bigelow, J. H. Zhang, W. M. Reiff, *J. Chem. Soc., Chem. Commun.* **1986**, 1026-1028.

[5] J. S. Miller, *J. Mater. Chem.* **2010**, *20*, 1846-1857.

[6] J. S. Miller, J. C. Calabrese, R. S. McLean, A. J. Epstein, *Adv. Mater. 4,* 498-501 (1992).

[7] J. S. Miller, A. J. Epstein, *J. Chem. Soc., Chem. Commun*. 1319-1325 (1998).

[8] K. I. Pokhodnya, M. Bonner, J.-H. Her, P. W. Stephens, J. S. Miller, *J. Am. Chem. Soc. 126*, 15592.

[9] J. Zhang, J. Ensling, V. Ksenofontov, P. Gütlich, A. J. Epstein, J. S. Miller, *Angew. Chem. Internat. Ed. 37*, 657-660 (1998).

[10]    K. H. Stone, P. W. Stephens, A. C. McConnell, E. Shurdha, K. I. Pokhodnya, J. S. Miller, *Adv. Mater. 22*, 2514-2519 (2010).

[11]    J.-H. Her, P. W. Stephens, K. I. Pokhodnya, M. Bonner, J. S. Miller, *Angew. Chem. Internat. Ed. 46,* 1521-1524 (2007).

[12]    J. S. Miller, *Polyhed*. *28*, 1596-1605 (2009).

[13]    A. V. Soudackov, A. L. Tchougréeff, I. A. Misurkin, *Theor. Chim. Acta* **1992,** *83*, 389.

[14]    A. M. Tokmachev, A. L. Tchougréeff*, Khim. Fiz.* **1999**, *18*, No. 1, 80 [in Russian].

[15]    A. V. Soudackov, A. L. Tchougréeff, I.A. Misurkin. *Zh. Fiz. Khim.* **1994,** *68*, 1256 [in Russian]; *Russ. J. Phys. Chem.* **1994,** *68*, 1135 [in English].

[16]    A. V. Soudackov, A. L. Tchougréeff, I.A. Misurkin. *Zh. Fiz. Khim.* **1994,** *68,* 1264 [in Russian]; *Russ. J. Phys. Chem.* **1994**, *68*, 1142 [in English].

[17]    A. V. Soudackov, A. L. Tchougréeff, I. A. Misurkin. *Int. J. Quant. Chem.* **1996,** *57*, 663.

[18]    A. V. Soudackov, A. L. Tchougréeff, I. A. Misurkin. *Int. J. Quant. Chem.* **1996,** *58*, 161.

[19]    A. L. Tchougréeff. *J. Mol. Catal.* **1997,** *119*, 377.

[20]    A. L. Tchougréeff. *Khim. Fiz.* **1998,** *17*, No 6, 163 [in Russian]; *Chem. Phys. Reports.* **1998,**

*17*, No 6, 1241 [in English].

[21] A. V. Sinitsky, M. B. Darkhovskii, A. L. Tchougréeff, I. A. Misurkin. *Int. J. Quant. Chem.* **2002,** *88*, 370.

[22] A. M. Tokmachev, A. L. Tchougréeff. *J. Sol. State Chem.* **2003,** *176*, 633.

[23] M. B. Darkhovskii, A. V. Soudackov, A. L. Tchougréeff. *Theor. Chem. Acc.* **2005,** *114*, 97.

[24] M. B. Darkhovskii, A. L. Tchougréeff. in Recent Advances in Theory of Chemical and Physical Systems J.-P. Julien, J. Maruani, and E. Brändas (eds), **2006**, Kluwer, Dordrecht.

[25] X.-H. Liu, R. Dronskowski, R. Glaum, A. L. Tchougréeff. *Z. Allg. & Anorg. Chem.* 636 (2010) 343-348.

[26] X.-H. Liu, M. Speldrich, P. Kögerler, R. Dronskowski, A. L. Tchougréeff. *Inorg. Chem.* **2010**, *49*, 7414.

[27] A. L. Tchougréeff, R. Dronskowski. *J. Phys. Chem. A* **2011**, *115*, 4547.

[28] J. B. Goodenough. Magnetism and the Chemical Bond. Interscience-Wiley, NY (1963). S. V. Vonsovskii, Magnetism. Nauka, Moscow (1971) [in Russian]; S. V. Vonsovsky, Magnetism. Wiley, NY (1974) in two volumes.

[29] T. Koopmans. *Physica* **1934**, *1*, 104; I. Mayer. Simple Theorems, Proofs and Derivations in Quantum Chemistry. Springer. **2003**, NY *et al*.

[30] A. L. Tchougréeff, R. Dronskowski. *J. Comp. Chem.*, **2008**, *29*, 2220; A. L. Tchougréeff, R. Dronskowski. *Int. J. Quant. Chem.* **2011**, *111*, 2490.

[31] R. D. Shannon. *Acta Cryst. A* **1976**, *32*, 751; R. Dronskowski. Computational Chemistry of Solid State materials. Wiley-VCH, Weinheim (2005).

[32] W. E. Buschmann, J. S. Miller. Chem. Eur. J. 4 (1998) 1731.

[33] S. F. Rach, F. E. Kühn. *Chem. Rev.* **2009,** *109,* 2061.

[34] G. C. De Fusco, L. Pisani, B. Montanari, N. M. Harrison. *Phys. Rev. B* **2009**, *79*, 085201.

Table 1: Quantities as necessary for estimating the exchange parameters in metal-TCNE solids from EHCF calculations on model complexes cut from respective crystals.

| | V(TCNE)$_2$ | Fe(TCNE)$_2$ | Mn(TCNE)$_2$ | Mn TCNE I$_3$ layer | Mn TCNE I$_3$ wall |
|---|---|---|---|---|---|
| $g_{dd}$ (eV) | 13.36 | 15.37 | 14.70 | 14.70 | 14.70 |
| $I^{(0)}_d$ (eV) | 21.10 | 21.02 | 21.02 | 20.93 | 20.93 |
| $\Delta E^{(0)}_{d \to LUMO}$ (eV) | 12.19 | 12.67 | 15.16 | 14.83 | 14.89 |
| $g_{dSOMO}$ (eV) | 3.08 | 3.15 | 3.09 | 3.10 | 3.13 |
| $\sum_\mu \beta^2_{\mu SOMO}$ [1] (eV$^2$) | 0.049 | 0.024 | 0.034 | 0.071 | 0.032 |
| $\varepsilon_{LUMO}$ (eV) | −5.83 | −5.20 | −5.86 | −6.10 | −6.04 |
| $\Delta E_{d \to SOMO}$ (eV) | 1.17 | 2.70 | 2.54 | 2.87 | 2.81 |
| $\Delta E_{SOMO \to d}$ (eV) | 9.11 | 9.52 | 9.07 | 8.73 | 8.76 |
| $J_{d \to SOMO}$ [2] (meV) | 12 | 6 | 8 | 18 | 8 |
| $J_{SOMO \to d}$ [2] (meV) | 84 | 18 | 27 | 49 | 23 |
| $E_{ex}$ (meV) | 96 | 24 | 35 | 67 | 31 |

[1] The sum over singly occupied $d$-orbitals μ is meant.
[2] The sums of the squared one-electron hopping integrals divided by the respective energy denominators are meant.

Table 2a. Exchange energy, $E^o_{ex}$, for the "wall" model of Mn-TCNE compound as function of $\angle MNC_{TCNE}$.

| $\angle MNC_{TCNE}$, (deg) | $E^0_{ex}$ (meV) |
|---|---|
| 173 | 8.6 |
| 158 | 13.81 |
| 143 | 27.92 |
| 128 | 47.02 |
| 113 | 65.33 |

Table 2b Exchange energies, $E^o_{ex}$ (meV), for the *"layer"* models $[M^{II}(NCMe)_5TCNE]^+$ as functions of bending angle and $M^{II}$

|  | V | Cr | Mn | Fe | Co | Ni | Cu |
|---|---|---|---|---|---|---|---|
| 0  | 15.94 | 9.45  | 19.07 | 7.94  | 0.16  | 0.17  | 0.05 |
| 15 | 14.94 | 11.39 | 33.76 | 9.53  | 3.44  | 3.84  | 1.43 |
| 30 | 11.63 | 19.34 | 48.96 | 20.95 | 15.64 | 17.33 | 6.37 |
| 45 | 6.87  | 30.08 | 58.94 | 38.34 | 32.83 | 36.42 | 13.28 |
| 60 | 2.06  | 12.44 | 64.69 | 55.19 | 48.26 | 53.63 | 19.09 |

Table 3: Exchange integrals' values for materials studied.

| Material | type | $n_{unp}$ | $J/k_B = E_{ex}/n$ |
|---|---|---|---|
| Mn(TCNE) $I_3$ | layer | 5 | 155 |
| Mn(TCNE) $I_3$ | wall | 5 | 72 |
| Fe(TCNE)$_2$ | layer | 4 | 70 |
| V(TCNE)$_2$ | layer | 3 | 370 |
| Mn(TCNE)$_2$ | layer | 5 | 80 |

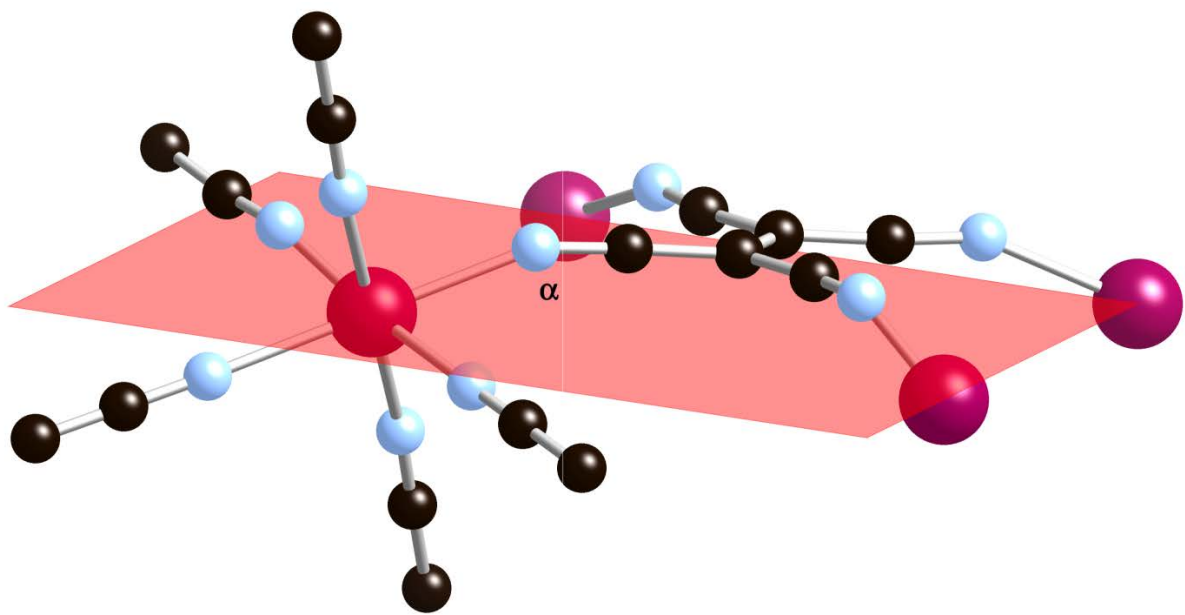

**Figure 1.** Structure of the $[M^{II}(NCMe)_5(TCNE)]^+$ moiety computationally studied. H atoms omitted for clarity. The ∠M-N-C valence angle (α) has been varied by up to 60° in order to estimate the geometry dependence of the exchange energy. Details are given in the text.

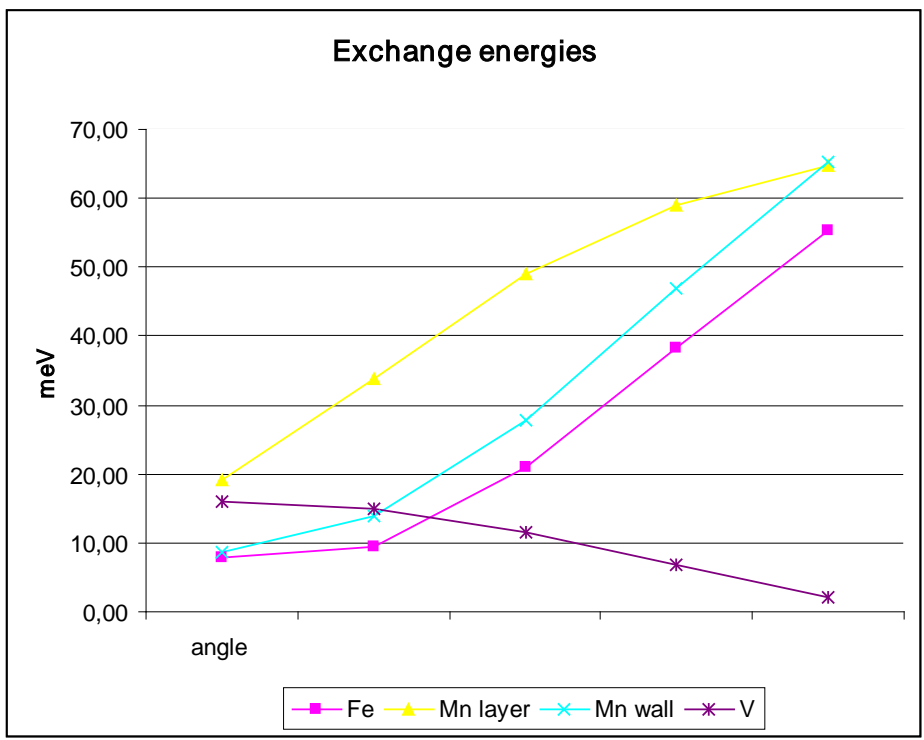

Figure 2. Exchange energies for the series of bending angles (degrees) 180 - ∠MNC of the four model clusters as listed in Table 2.